\documentclass{pasj00}
\usepackage{times}
\usepackage{mathptm}

\begin{document}
\SetRunningHead{H. Nakanishi, et al. }{ASTE $^{12}$CO($J=$3--2) Survey of Elliptical Galaxies}
\Received{2006/09/29}
\Accepted{2006/11/17}

\title{ASTE $^{12}$CO($J=$3--2) Survey of Elliptical Galaxies} 

\author{Hiroyuki \textsc{Nakanishi}\altaffilmark{1}, Tomoka \textsc{Tosaki}\altaffilmark{1}, Kotaro \textsc{Kohno}\altaffilmark{2}, Yoshiaki \textsc{Sofue}\altaffilmark{2}, and Nario \textsc{Kuno}\altaffilmark{1}}%
\email{hnakanis@nro.nao.ac.jp}
\altaffiltext{1}{Nobeyama Radio Observatory, Minamimaki, Minamisaku, Nagano 384-1305}
\altaffiltext{2}{Institute of Astronomy, The University of Tokyo, 2-21-1 Osawa, Mitaka,Tokyo 181-0015}

\maketitle
\begin{abstract}
We report $^{12}$CO($J=$3--2) observations of 15 nearby elliptical galaxies carried out with the ASTE telescope. Thirteen were selected without regard to the presence of other tracers of cold interstellar matter. CO emission was detected from three of the galaxies, two of which are undetected by IRAS at 100 microns, suggesting that cold ISM may be present in more ellipticals than previously thought. The molecular gas masses range from $2.2 \times 10^6$ to $4.3 \times 10^8$ $M_\odot$. The ratio of the CO(3--2) and (1--0) lines, $R_{31}$, has a lower value for elliptical galaxies than for spiral galaxies except for NGC 855, for which the value is close to the mean for spirals. {The molecular gas in NGC 855 has a mean density in the range 300 -- 1000 cm$^{-3}$ adopting a temperature range of 15 -- 100 K.}
\end{abstract}

\section{Introduction}
Spiral galaxies contain substantial amounts of cold interstellar matter (ISM) and ongoing star formation, while cold ISM is but a minor component of elliptical galaxies. This is generally attributed to early exhaustion of the ISM in ellipticals during the early stages of galaxy formation. However, mass loss during stellar evolution \citep{fab76} can replenish the ISM to detectable levels in much less than a Hubble time, and ISM present in ellipticals can be attributed to stellar mass loss or to accretion from outside.

The mass loss from evolved stars is estimated to be $1.5 M_\odot$ per year per $10^{11} L_\odot$ of optical luminosity \citep{fab76} and indeed is observed to take place as cold molecular gas (e.g., \cite{tey06}). 

Thus, ISM contents comparable to those of spiral galaxies could be accumulated in elliptical galaxies in of order $10^9$ years. However, \citet{kna85} point out that the cold gas and stellar contents of elliptical galaxies are uncorrelated, and conclude that the cold gas comes from outside through events such as mergers. On the other hand, many elliptical galaxies contain of order $10^8$ to $10^{10} M_\odot$ of hot gas, detected by its X-ray emission \citep{for85}, which could originate in collisionally-heated mass shed by dying stars (e.g., \cite{mat03}). 

Knowledge of the cold-gas content of normal elliptical galaxies is crucial to these questions. Several such surveys have been made \citep{lee91,sof93,wik95,kna96} using the CO(1--0) and CO(2--1) rotational lines as probes. However, all of these studies have concentrated on ellipticals which are detected at 100 microns by IRAS, and are therefore biased. The present paper discusses observations of 15 nearby elliptical galaxies selected independently of their infrared emission. These observations were made of the CO(3--2) line, which in addition to providing information on the presence of molecular gas allows an evaluation of its physical condition from multi-transition measurements (cf. \cite{li93,vil03}).

\section{Observations}
We carried out $^{12}$CO($J=$3--2) (hereafter CO(3--2)) observations of fifteen elliptical galaxies using the Atacama Submillimeter Telescope Experiment (ASTE), a 10m antenna located at Pampa La Bola in the Atacama Desert in Chile at an altitude of 4800 m \citep{eza04,koh05}. The observations were made on August 21--24 2006 from the remote ASTE operations room at the Nobeyama Radio Observatory (NRO) using the network operation system N-COSMOS 3 developed at the National Astronomical Observatory of Japan (NAOJ) \citep{kam05}. The front end is a cartridge-type 350 GHz receiver operated in double-sideband mode with an intermediate frequency (IF) between 5 and 7 GHz. The antenna temperature and sky extinction were measured using the standard chopper-wheel method. The antenna beam size is $\timeform{22''}$ at 345 GHz. We used three digital spectrometers with a bandwidth of 512 MHz (445 km s$^{-1}$ at 345 GHz) and 1024 channels. {The same center frequency was set for these spectrometers. A velocity resolution was 0.53 km s$^{-1}$.} The pointing was monitored by observations of the bright line sources AFGL 3068 and $o$ Ceti, and was accurate to 5". The main beam efficiency $\eta_{\rm MB}$ was measured each day by observations of W51D using the peak intensity measured by \citet{wan94} and was constant at 0.54.

The observed galaxies were selected from the catalogue of \citet{kna89} to lie in the right ascension range 22h - 7h and between declinations of $-30$ degrees -- $+20$ degrees (so that the galaxies can be observed from both the Northern and Southern hemispheres), to have recession velocities less than 5000 km s$^{-1}$ and to be classified as ellipticals rather than as S0s or of later types. No other criterion was used for thirteen of the galaxies, so they are unbiased. In addition, two infrared-bright dwarf ellipticals, NGC 855 and NGC 2328, were observed. In these galaxies, CO(1--0) and CO(2--1) emission have been observed and the line widths are less than 445 km s$^{-1}$. The observed galaxies are listed in table 1.

The data reduction was carried out with tasks in NEWSTAR, a package developed at NRO. Individual bad data were flagged, the remaining spectra averaged, and first-order baselines were removed. The calibrated spectra were smoothed by 46 channels to a velocity resolution of 19.8 km s$^{-1}$. The typical rms noise levels, $\Delta T_{\rm MB}$, for the smoothed spectra were 8 mK.

\begin{table*}[h]
 \begin{center}
 \caption{Target list}\label{coefficient}
  \begin{tabular}{cccccccccc}
   \hline\hline
   Name&\multicolumn{3}{c}{R.A.(B1950)} &\multicolumn{3}{c}{Dec.(B1950)}
   &Morph.& B$_T$& $V_{\rm HEL-OPT}$\\
   & h & m & s & d&m&s&&mag.&km s$^{-1}$\\
   \hline
	\multicolumn{10}{c}{First samples}\\
\hline
 NGC596  & 01 & 30 & 22 & $-07$ & 17 & 18 & E1 & 11.8 & 1817 \\
 NGC636  & 01 & 36 & 36 & $-07$ & 45 & 54 & E3 & 12.4 & 1805 \\
 NGC821  & 02 & 05 & 41 & $+10$ & 45 & 32 & E6 & 11.7 & 1716 \\
 NGC990  & 02 & 33 & 36 & $+11$ & 25 & 28 & E  & 13.9 & 3508 \\
 NGC1453 & 03 & 43 & 57 & $-04$ & 07 & 36 & E2 & 12.6 & 3906 \\
 NGC1550 & 04 & 17 & 02 & $+02$ & 17 & 25 & E  & 14.0 & 3689 \\
 NGC1600 & 04 & 29 & 12 & $-05$ & 11 & 30 & E4 & 12.0 & 4687 \\
 NGC7458 & 22 & 58 & 55 & $+01$ & 29 & 05 & E  & 13.9 & 4981 \\
 NGC7464 & 22 & 59 & 25 & $+15$ & 42 & 17 & E1p& ---  & 1877 \\
 NGC7468 & 23 & 00 & 30 & $+16$ & 20 & 08 & E3p& 14.0 & 2089 \\
 NGC7619 & 23 & 17 & 43 & $+07$ & 55 & 57 & E2 & 12.3 & 3747 \\
 NGC7626 & 23 & 18 & 10 & $+07$ & 56 & 35 & E  & 12.3 & 3450 \\
 NGC7785 & 23 & 52 & 45 & $+05$ & 38 & 11 & E5 & 12.6 & 3826 \\ 
\hline
	\multicolumn{10}{c}{Second samples}\\
\hline
 NGC855  & 02 & 11 & 10 & $+27$ & 38 & 36 & E  & 13.3 & 600 \\
 NGC2328 & 07 & 01 & 01 & $-41$ & 59 & 42 & E/S0 & 13.3  & 1159 \\
   \hline
\multicolumn{10}{l}{These parameters are taken from \citet{kna89}.}\\
  \end{tabular}\\
 \end{center}
\end{table*}

\section{Results}
\subsection{Spectra}
\begin{figure*}[h]
  \begin{center}
    \rotatebox{-90}{\FigureFile(120mm,120mm){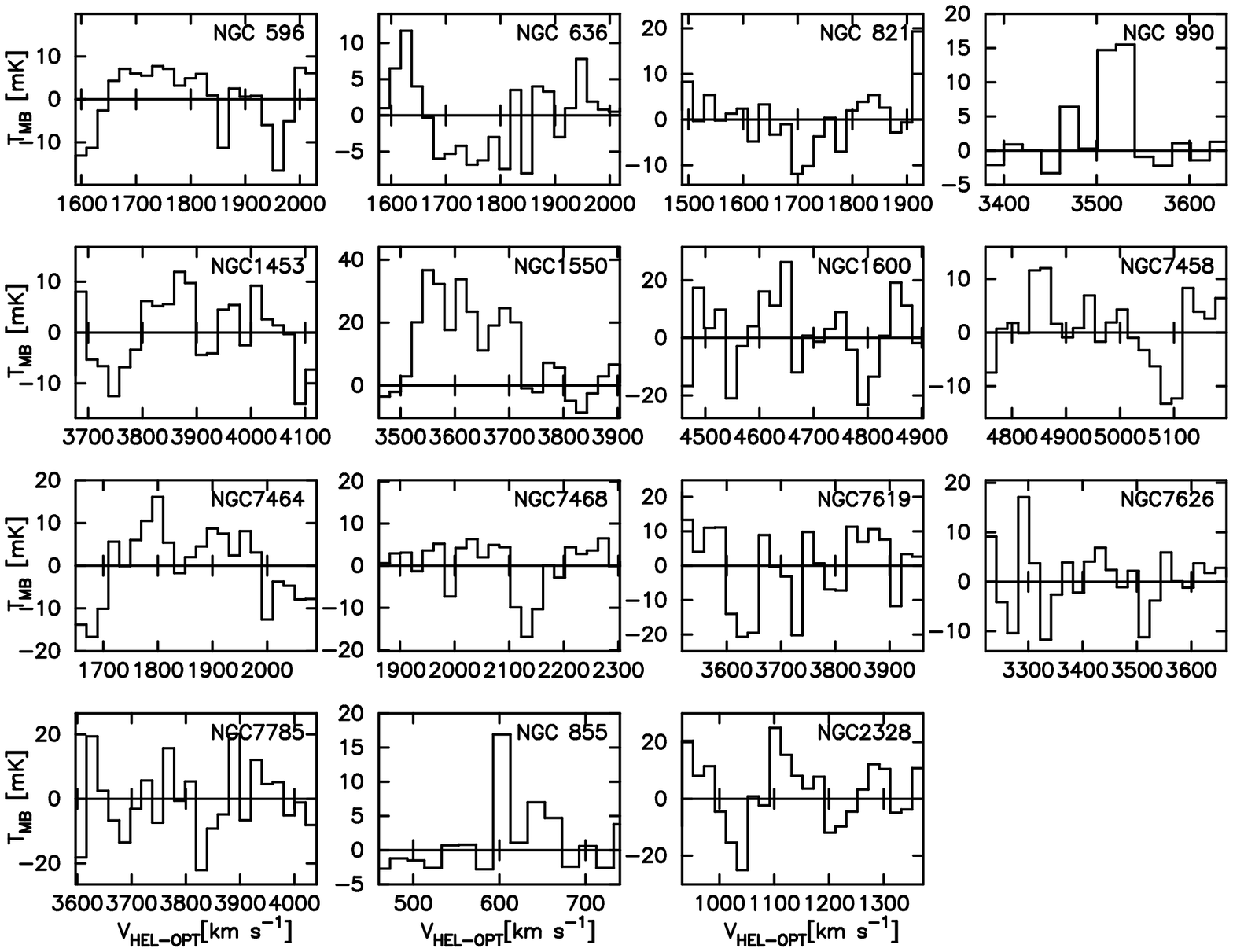}}
  \end{center}
  \caption{CO Spectra of all the observed elliptical galaxies. \label{spectra}}
\end{figure*}

The spectra for all fifteen observed galaxies are shown in figure \ref{spectra} and the observational quantities are listed in table 2. We detected CO(3--2) emission from three galaxies, all for the first time (the observation of NGC 855 by \citet{vil03} is not sufficiently sensitive to detect emission at the level of that measured in this paper). 

\begin{figure*}[h]
  \begin{center}
    \rotatebox{0}{\FigureFile(90mm,100mm){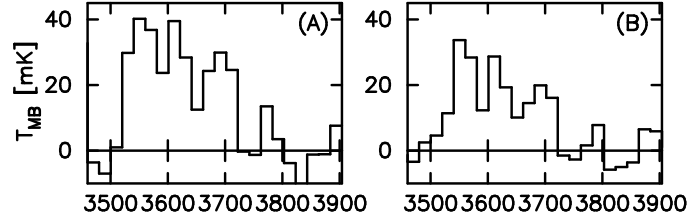}}
  \end{center}
  \caption{Spectra for NGC 1550. We halved the original data into two independent groups. Presented spectra are obtained by averaging the spectra belonging to the individual groups. \label{sp-n1550}}
\end{figure*}

{The spectrum for NGC 1550 is highly asymmetric relative to the systemic velocity measured in optical observations.} In order to examine whether this detection is real, we halved the original data into two independent groups, and then averaged the spectra belonging to each group. In the both cases, similar profiles were obtained as shown in figure \ref{sp-n1550}. {Hence, we can conclude that this detection is likely to be real. }

{Although IRAS 100 micron flux in NGC 7464 is the highest in our observed sample, CO emission has not been detected \citep{wik95}.} NGC 7468 was detected in only CO(1--0) line at $\timeform{22''}$ beam \citep{wik95}, which is the same as our observing beam size. {Since the CO(1--0) peak temperature is about 20 mK, the CO(3--2) peak temperature would be about 4 mK, assuming the mean line ratio for elliptical galaxies of $R_{31} = 0.21$ \citep{vil03}.} The rms noise of 7 mK in our observation was not low enough to detect CO(3--2) emission. 

{Although NGC 2328 was detected in both CO(1--0) and (2--1) lines \citep{wik95,lee91}, we could not detect CO(3--2) emission. The peak temperature of the CO(1--0) line was about 15 mK at $\timeform{44''}$ beam. If the size of the molecular distribution was less than $\timeform{22''}$, and if the $R_{31}$ was 0.21, the CO(3--2) peak temperature would be about 13 mK. The rms noise in our observation was too high (12 mK) to detect the line. }

{Note that CO(3--2) emission could not be detected if a velocity width was larger than the bandwidth of the spectrometers (445 km s$^{-1}$). }

\subsection{Molecular Masses}
We assume the average ratio $R_{31}$ of 0.21 to estimate the expected CO(1--0) flux for two of the galaxies, and use the CO(1--0) data for NGC 855 from \citet{wik95}. The molecular hydrogen mass is calculated from
\begin{equation}
M_{\rm H2} = {\pi \over 4} X_{\rm CO} {I_{{\rm CO}(3-2)}\over R_{31}} m_{\rm H2} \theta^2 d^2. 
\end{equation} 
The distance was calculated from the recession velocity $V$ assuming a Hubble constant of 72 km s$^{-1}$ Mpc$^{-1}$ \citep{spe03}. The conversion factor was taken to be $X_{\rm CO} = 2.8 \times 10^{20}$ cm$^{-2}$ K$^{-1}$ km$^{-1}$ s, an average of the values found for nearby Galactic molecular clouds by \citet{ari96}. {The resulting masses are listed in table 3 and cover a similar range to those measured for other elliptical galaxies \citep{lee91,sof93,wik95}.}

\begin{table*}[h]
 \begin{center}
 \caption{Observational quantities}\label{obs-quantities}
  \begin{tabular}{cccccccc}
   \hline\hline
   Name& $F_{{100\mu}{\rm m}}$& $V_{\rm CO}$ & $\Delta V_{\rm CO}$& $T_{\rm peak}$ & $I_{\rm CO}$ & $\Delta T_{\rm MB}$ & comment \\
       & Jy & km s$^{-1}$  & km s$^{-1}$         &  mK            & K km s$^{-1}$& mK  &         \\
   (1) & (2) & (3) & (4) & (5) & (6) & (7) & (8)\\
   \hline
	\multicolumn{8}{c}{First samples}\\
\hline
 NGC596  &$\le 0.300$& ---  & --- &   8 & ---  &  8 & no detection \\ 
 NGC636  &$\le 0.399$& ---  & --- &  12 & ---  &  5 & no detection \\ 
 NGC821  &  0.440 & ---  & --- &  19 & ---  &  6 & no detection \\
 NGC990  &$\le 2.24$& 3531 &  40 &  16 & 0.59 &  3 & detection \\
 NGC1453 &  0.670 & ---  & --- &  12 & ---  &  7 & no detection \\
 NGC1550 &$\le 0.651$& 3631 &  198 &  37 & 4.9  &  5 & detection \\
 NGC1600 &  0.170 & ---  & --- &  26 & ---  & 15 & no detection \\ 
 NGC7458 &$\le 0.192$& ---  & --- &  12 & ---  &  8 & no detection \\
 NGC7464 & 6.71 & ---  & --- &  16 & ---  &  8 & no detection \\
 NGC7468 & 1.65 & ---  & --- &   7 & ---  &  7 & no detection \\
 NGC7619 & 0.630 & ---  & --- &  13 & ---  & 11 & no detection \\
 NGC7626 &$\le 0.339$ & ---  & --- &  17 & ---  &  7 & no detection \\
 NGC7785 &$\le 0.264$& ---  & --- &  20 & ---  & 11 & no detection \\
\hline
	\multicolumn{8}{c}{Second samples}\\
\hline
 NGC855  & 3.27 & 613  &  20 &  17 & 0.36 & 4  & detection \\
 NGC2328 & 3.77 & ---  & --- &  25 & ---  & 12 & no detection \\
\hline
\multicolumn{8}{l}{Col. 1.--- Galaxy name.} \\
\multicolumn{8}{l}{Col. 2.--- IRAS $100\mu$m flux. The upper limit was taken to be $3\sigma$.} \\
\multicolumn{8}{l}{Col. 3.--- Central velocity.} \\
\multicolumn{8}{l}{Col. 4.--- Velocity width (Full width at zero intensity; FWZI).} \\
\multicolumn{8}{l}{Col. 5.--- Peak brightness temperature.} \\
\multicolumn{8}{l}{Col. 6.--- Integrated intensity.} \\
\multicolumn{8}{l}{Col. 7.--- Rms noise.}\\
\multicolumn{8}{l}{Col. 8.--- Comment on detection.} \\
  \end{tabular}
\end{center}

\end{table*}

\begin{table*}[h]
 \begin{center}
 \caption{Inferred Properties of Molecular Gas}\label{mass-etc}
  \begin{tabular}{lccc}
   \hline\hline
   Name& $M_{\rm H2}$ & Volume  & Diameter \\
       & $M_\odot$    & pc$^{3}$&  pc  \\
   \hline
 NGC990  & $2.7\times 10^8$ & ($8.3\times 10^6$) & (250) \\
 NGC1550 & $4.3\times 10^8$ & ($1.3\times 10^7$) & (300) \\ 
 NGC855  & $2.2\times 10^6$ & $6.9\times 10^4$ & (51) \\ 
\hline
  \end{tabular}
\end{center}

\end{table*}

\section{Discussion}
\subsection{Line Ratios for NGC 855}
Both CO(1--0) and CO(2--1) emission have been detected from NGC 855 \citep{lee91,wik95}. Comparison with the observations of \citet{wik95} gives line ratios of $R_{31} = 0.5$ and $R_{21} = 0.8$. The beam sizes used for the CO(3--2) and CO(1--0) are the same, 22", while the beam size for the CO(2--1) observation is $\timeform{11''}$. Therefore, the true value of $R_{21}$ might be a factor 1 to 4 lower, while $R_{31}$ can be considered as the true value. 

The value of $R_{31}$ is larger than the mean observed for three other elliptical galaxies (0.2, \cite{vil03}), but similar to values observed for spiral and starburst galaxies (Mauersberger et al. 1999).

\subsubsection{Star Formation Efficiency}
The star formation efficiency (SFE) of NGC 855 is about $8 \times 10^{-10}$ yr$^{-1}$ \citep{vil03}, higher than that of most elliptical galaxies and similar to values observed for spiral galaxies \citep{kom05}. {This suggests that $R_{31}$ correlates with star formation efficiency, as noted from mapping of M83 by \citet{mur06a} and \citet{mur06b}.}

\subsubsection{Modeling the Molecular ISM}
Since we have data for three CO rotational in NGC 855, the mean properties of the molecular ISM can be modelled. First, we assume LTE and that all three CO lines are optically thick. The brightness temperature, $T_{\rm b}$, is then 
\begin{equation}
T_{\rm b} = {h\nu \over k} \left( {\left(\exp{\left({h\nu\over kT}\right)}-1\right)^{-1}}-{\left(\exp{\left({h\nu\over 2.7k}\right)}-1\right)^{-1}}\right), 
\end{equation} 
where $h$ is Planck's constant, $\nu$ is frequency, $k$ is Boltzmann's constant
and $T$ is the kinetic temperature. Applying equation (2) to the observed line ratios for NGC 855 implies a kinetic temperature of 8 K, which is far lower than the dust temperature of 33 K estimated by \citet{wik95}.

\begin{figure*}[h]
  \begin{center}
    \rotatebox{0}{\FigureFile(90mm,100mm){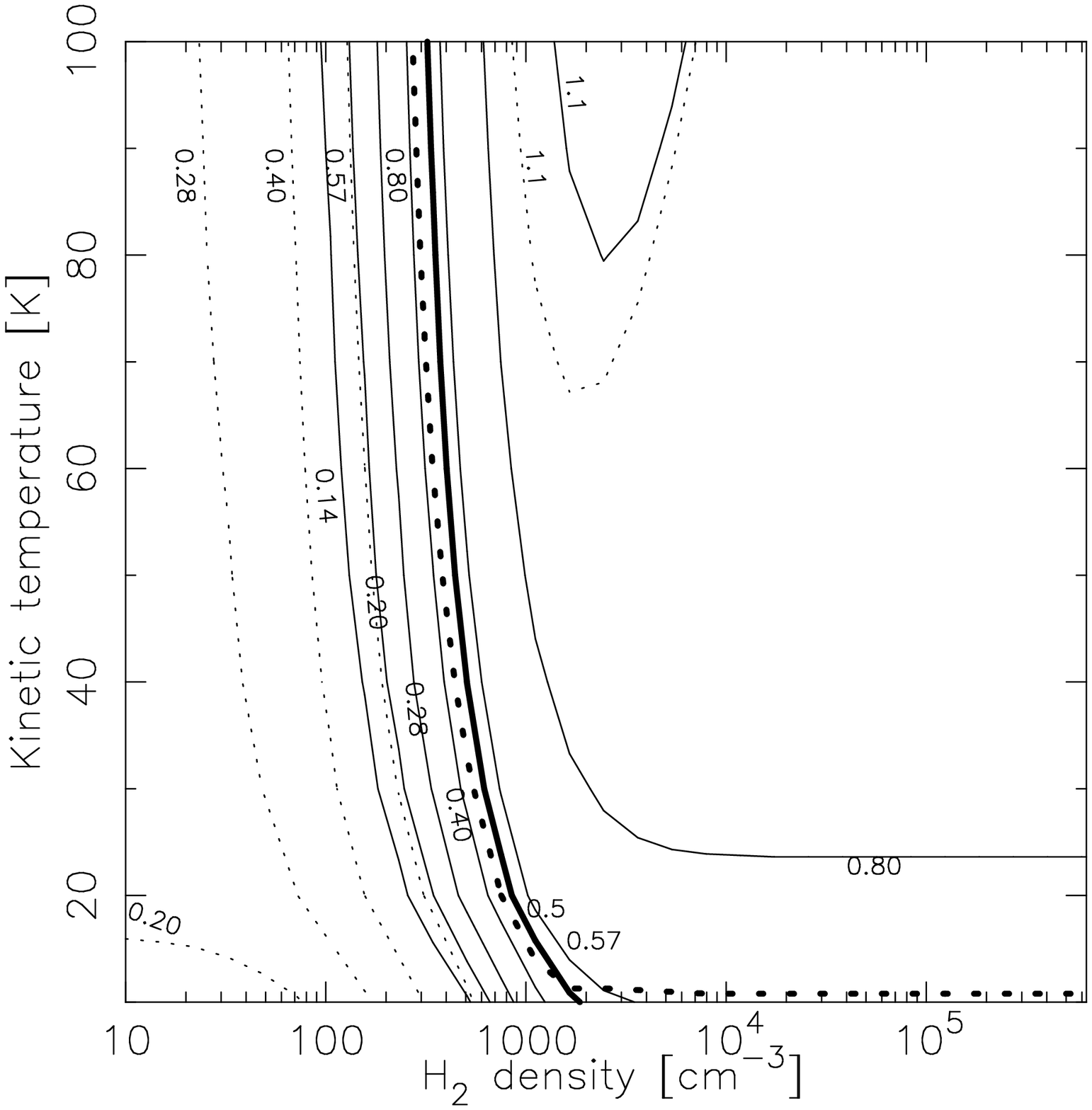}}
  \end{center}
  \caption{Results of LVG calculations adopting CO abundance of $X_{\rm CO} /(dV/dr) = 10^{-5}$ pc km$^{-1}$ s. The solid and dotted lines denote $R_{31}$ and $R_{21}$, respectively. The thick solid and dotted lines denote  $R_{31}$=0.5 and $R_{21}$=0.8, respectively. \label{LVG}}
\end{figure*}

Accordingly, we modelled the emission using the Large Velocity Gradient (LVG) model \citep{gol74}. We calculated line ratios for kinetic temperatures from 10 -- 100 K, H$_2$ densities from 10 to $6.3 \times 10^6$ cm$^{-3}$, and CO abundances $X_{\rm CO} (dV/dr)^{-1}$ of $10^{-6}$, $10^{-5}$ and $10^{-4}$ pc km$^{-1}$ s. The resulting line ratios $R_{31}$ and $R_{21}$ are shown in figure \ref{LVG}. The data and models match well for $T >$ 15 K, compatible with the dust temperature. The corresponding density is in the range of 300 -- 1000 cm$^{-3}$.  The CO(1--0) line is optically thick at this density, as was assumed in estimating the molecular mass from Equation (1).

\subsection{Volume and Size of Molecular Distribution}
Using the calculated mean density, we can estimate the volume of the molecular gas and its diameter, assuming a spherical distribution. 
We used a mean H$_2$ density of 650 cm$^{-3}$  for all three detected elliptical galaxies. The results are given in table 3. 
{The molecular content in NGC 855 is as much as a single Galactic Giant Molecular Cloud (GMC), which can support the existence of the star formation in this galaxy (e.g., \cite{vil03}). 

However, the size is smaller than the beam size of CO(2--1) ($\timeform{11''}$), which corresponds to 440 pc at 8.3 Mpc. Therefore, $R_{21}$ would be 4 times smaller (i.e., $R_{21}=0.2$), if NGC 855 had only one GMC. 
The LVG calculation cannot satisfy $R_{31}=0.5$ and $R_{21}=0.2$ at the same time, if this was the case. The curve of $R_{31}=0.5$ always coincides with that of $R_{21} \sim 0.8$. 

Considering these matters, the molecular distribution is likely to be more extended than the beam size $\timeform{22''}$, forming smaller clumps rather than one GMC.} 



\subsection{Comparison with IRAS fluxes}
Two of the elliptical galaxies (NGC 990 and NGC 1550) in which we detected molecular gas are not detected by IRAS, {though IRAS 100$\mu$m upper limit of NGC 990 is pretty high}. Thus cold molecular ISM may be more common in ellipticals than previously thought.

\section{Summary}
We achieved $^{12}$CO($J=$3--2) line survey of 15 elliptical galaxies and detected confident CO emissions from three ellipticals. 
The molecular gas masses were estimated to be $2.2\times10^6$ -- $4.3\times10^8$ $M_\odot$. 

Line ratio, $R_{31}$, of NGC 855 was calculated to be 0.5. 

We made line ratio analysis using $R_{31}$ and $R_{21}$ of NGC 855. 
The LVG model showed that H$_2$ density was $3\times 10^2$ -- $1\times 10^3$ cm$^{-3}$ in the temperature range of 15 -- 100 K. 

CO($J=$3--2) emission was detected even in elliptical with faint IRAS 100 $\mu$m flux. More ellipticals might be detected in the CO lines than previously thought.

\vspace{0.3cm}
We are grateful to all the ASTE staff for their dedicated support for our observation. We thank the referee Dr. Knapp for substantial corrections of English as well as invaluable scientific comments, which greatly improved our paper. 
We thank Dr. T. Sawada and Dr. S. Takano for fruitful discussions and advices for the observations. 
This study was financially supported by MEXT Grant-in-Aid for Scientific Research on Priority Areas No. 15071202.

\end{document}